\documentclass[prb,twocolumn,showpacs,preprintnumbers,amsmath,amssymb]{revtex4}

\usepackage{amsfonts}
\usepackage{latexsym}
\usepackage{graphicx}
\usepackage{dsfont}
\usepackage{epsfig}

\newcommand{\Eqref}[1]{Eq.~(\ref{#1})}

\newcommand{\be}{\begin{equation}}
\newcommand{\ee}{\end{equation}}

\begin{document}

\title{Theory of heating of hot magnetized plasma by Alfv\'en
waves.\\
Application for solar corona}
\author{T.M.~Mishonov}
\email[E-mail: ]{mishonov@phys.uni-sofia.bg}
\author{M.V.~Stoev}
\email[E-mail: ]{martin.stoev@gmail.com}
\author{Y.G.~Maneva}
\email[E-mail: ]{yanamaneva@gmail.com}
\affiliation{Department of Theoretical Physics, Faculty of Physics,\\
University of Sofia St.~Clement of Ohrid,\\
5 J. Bourchier Boulevard, BG-1164 Sofia, Bulgaria}

\date{\today}

\begin{abstract}
The heating of magnetized plasma by propagation of Alfv\'en waves is
calculated as a function of the magnetic field spectral density. The
results can be applied to evaluate the heating power of the solar
corona at known data from satellites' magnetometers. This heating
rate can be incorporated in global models for heating of the solar
corona and creation of the solar wind. The final formula for the
heating power is illustrated with a model spectral density of the
magnetic field obtained by analysis of the Voyager 1 mission
results. The influence of high frequency dissipative modes is also
taken into account and it is concluded that for evaluation of the
total coronal heating it is necessary to know the spectral density
of the fluctuating component of the magnetic field up to the
frequency of electron-proton collisions.
\end{abstract}

\pacs{52.35.Bj, 52.50.Sw, 96.60.P-, 94.30.cq}

\maketitle

%%%%%%%%%%%%%%%%%%%%%%%%%%%%%%%%%%%%%%%%%%%%%%%%%%%%%%%%%%%%%%%%%%%%
\section{Introduction}
%%%%%%%%%%%%%%%%%%%%%%%%%%%%%%%%%%%%%%%%%%%%%%%%%%%%%%%%%%%%%%%%%%%%

The discovery of the spectral lines of highly ionized iron in the
spectrum of the solar corona showed in 1939 that the corona is much
hotter than the radiating surface of the Sun. Why it is hotter by
about two orders of magnitude is still a very discussed problem,
which survived not only the Second World War, but also the whole
development of the modern astrophysics and satellite technologies in
the past half century. Definitely, this is the oldest still living
mystery in front of the contemporary physics.

On the other hand, heating of the solar corona is a phenomenon in
the simplest physical system - a hot ideal gas of electrons and
protons. All the phenomena in a completely ionized hydrogen plasma
are well-known and now it is a matter of a carefully proposed
complementary set of experiments to arrive at the final conclusion.
According to the second law of thermodynamics each process in the
plasma surrounding the Sun produces some entropy and can be
considered as a heating mechanism of the solar corona. At least one
new idea per year appears under the Sun, concerning the heating of
its corona. In a review article on the subject one can find more
than a hundred references. Often they are being classified as $AC$
and $DC$ models \cite{Mandrini00} depending on whether a direct or
an alternating electric current heats the plasma. But the coronal
plasma can not be heated by an electric current, as it was proved
thirty years ago \cite{Feldman}: ``...it is difficult to understand
how a cooler gas (the electrons) can heat a hotter gas (the protons)
within a region so far away from the Sun, which is the ultimate
source of most of the energy in interplanetary space'', see
\cite{Schwenn}. As far as the proton temperature is higher than the
electron one $T_p>T_e$ the heating mechanism acts on the protons and
the viscose friction of the shear flow with a velocity field
$\mathbf{V}(z,t)$ is a natural explanation for the bulk power $Q$ of
the dissipated energy \cite{LL06}
\be Q=\frac{1}{2}\eta \sum_{i,j=1,2,3} (\delta_i V_j+\delta_j
V_i)^2, \ee
for $\mathrm{div} V=0$. Such a heating process can be observed in
every turbulent fluid. Imagine a waterfall for which the amplitude
of turbulent pulsations is smaller than the velocity of sound and as
a result the related heating after the calming of the turbulent
motion is smaller $\Delta T/T\ll1$. That is why the turbulence in
the solar plasma cannot give an augmentation by two orders of
magnitude of the coronal temperature. For an effective heating
mechanism we need an effective mechanism for the energy transport.

This idea is supported by the recent development of the solar
physics related to the random motion of magnetic footprints. The
creation of Alfv\'en waves by the stochastic convection in the Sun
is a very interesting hydrodynamic problem for the complete heating
scenario (for additional references see Ref.~3). The purpose of the
present work is to analyze what kind of local satellites'
measurements should be done in order to confirm or reject the
Alfv\'en wave mechanism for the coronal heating. Nowadays
astrophysicists are able to use the available data provided by SOHO,
TRACE, ACE, WIND in order to probe their model. However, the main
defect of these experiments is the uncompleted set of data and that
is why as far as the theory is concerned all the gladiators remain
friendly sleeping in the arena. In the next section we will
systematize the formulas for Alfv\'en waves from well-known
textbooks and in the final section we will analyze what kind of
research has to be done.

%%%%%%%%%%%%%%%%%%%%%%%%%%%%%%%%%%%%%%%%%%%%%%%%%%%%%%%%%%%%%%%%%%%%
\section{Basic notions. Model evaluation}
%%%%%%%%%%%%%%%%%%%%%%%%%%%%%%%%%%%%%%%%%%%%%%%%%%%%%%%%%%%%%%%%%%%%

Calculation of the dissipated power density is an indispensable
ingredient in every theory which claims to reveal the coronal
heating mechanism. The real scientific problem is to explain the
heating power, not the local temperature jump. The local temperature
is only a parameter to be used in a realistic global model for the
solar corona, which includes the acceleration of the solar wind by
an expansion of the heated plasma and all processes for the local
energy balance.

We choose the z-axis oriented by the constant external magnetic
field $\mathbf{B}_0=\mathbf{e}_z.$ We will investigate an Alfv\'en
wave, propagating on this axis. We can choose the y-axis to be
perpendicular to the external magnetic field and the wave vector
$\mathbf{k}=k_x\mathbf{e}_x+k_z\mathbf{e}_z$. The Alfv\'en waves are
linearly polarized and the wave component of the magnetic field
$\mathbf B = \mathbf{B}_0 + \mathbf{B}^{'}$ is oriented along the
y-axis
\be \mathbf{B}^{'}(\mathbf{r},t)=B_y^{'}\mathbf{e}_y, \quad
B_y^{'}=B_{y,0} \cos(\mathbf{k}\cdot\mathbf{r}-\omega_A t)e^{-\gamma
z},\ee
where the null index denotes the wave amplitude, $\omega_{A}$ is the
frequency of the Alfv\'en waves and the amplitude decrement $\gamma$
is supposed to be small enough. For the wave period averaged density
of the magnetic energy we have
\be w_B(z)=\frac{B_{y,0}^2}{2 \mu_0} \frac{1}{2} e^{-2 \gamma z},
\ee
where $\mu_0=4\pi$ or $4\pi*10^{-7}$ in the practical system.

If we follow the evolution of a wave-packet, the density of the
total energy (magnetic and kinetic) decreases exponentially over
time
\be w(t)=\frac{B_{y,0}^2}{2 \mu_0}e^{-t/\tau_A}, \ee
where the absorption rate of the Alfv\'en waves
\be
\frac{1}{\tau_A}=(\nu_\mathrm{kin}+\nu_{\mathrm{mag}})\mathbf{k}^2,
\quad \nu_\textrm{kin}=\frac{\eta}{\rho}, \quad
\nu_{\mathrm{mag}}=c^2\varepsilon_0/\sigma \ee
is given by a sum of the kinetic $\nu_\mathrm{kin}$ and the magnetic
$\nu_{\mathrm{mag}}$ viscosities, expressed in terms of the plasma
viscosity $\eta,$ density $\rho$ and the electrical conductivity of
the fluid $\sigma$; $\varepsilon_0=1/4\pi$ or
$\varepsilon_0=1/\mu_0c^2$ in the practical system.

For an order of magnitude evaluation of the amplitude decrement
\cite{LLproblem} we will consider only the Alfv\'en waves
$\omega_A=V_Ak_z$ propagating along the external magnetic field
lines $\mathbf{k}=k\mathbf{e_z}$ and discard the negligible
influence of the resistivity $\nu_{\mathrm{mag}}$
\be \frac{1}{\tau_A}=2 \gamma
V_A=\frac{\omega_A^2}{V_A^2}\frac{\eta}{\rho}=\frac{(2 \pi
f)^2}{V_A^2}\nu_\mathrm{kin}, \ee
where $f=\omega/2\pi$ is the circular frequency. The relation
between the velocity of the Alfv\'en waves and the magnetic field
can be found in any textbook on magnetohydrodynamics
\be \frac{1}{2} \rho V_A^2=\frac{1}{2 \mu_0} B_0^2, \quad
\mathbf{V}_A=\frac{\mathbf{B}_0}{\sqrt{\mu_0\rho}}, \quad
\mathbf{B}_0=B_0\mathbf{e}_z. \ee
We presume small fluctuations from the constant magnetic field so
that the maximal distortion angle of the magnetic field lines
\be b_{y,0}=\frac{B_{y,0}}{B_0} \ll 1 \ee
is also small and we can use the linear theory of Alfv\'en waves,
propagating along the z-axis, whose phase velocity $V_A$ coincides
with their group one $v_{gr}$
\be \omega_A(\mathbf{k})=\left|\mathbf{V}_A\cdot\mathbf{k}\right|,
\;\; \omega_A=V_A k_z, \;\; k_z>0, \;\,v_{gr}=V_A.  \ee
For the bulk power density of the dissipated wave energy we obtain
\be Q=\frac{w}{\tau_A}=\frac{B_{y,0}^2}{2 \mu_0 \tau_A}. \ee
In this model evaluation we have formally taken $\mathbf r =0,$
assuming that the weak space dependence of the magnetic field is
included in the local value of the wave amplitude,
\be B^{'}_y(t)=B_{y,0}^{(\omega)} \cos(\omega t), \ee
measured by an AC satellite's magnetometer. Here the upper index
$(\omega)$ emphasizes that up to now we have considered a
monochromatic wave whereas for the calculation of the total
absorption rate we have to perform a frequency summation over the
whole Alfv\'en waves spectrum. In reality we ought to take into
account also the influence of the Doppler effect, caused by the
solar wind, but as a gedanken-experiment let us consider, that the
frequency is measured in a system, bound to the coronal plasma.
Imagine that the magnetometer has a frequency-filter and detects
only frequencies in a narrow range $(f_1,f_2),$ where $f_2-f_1\ll
f_1$. The most important characteristic of the Alfv\'en waves is
their spectral density \cite{pictures}, given by a sum of the wave
energies in a narrow frequency interval
\be \Phi(f)=\frac{1}{f_2-f_1}\sum_{f \in (f_2,\;f_1)}
\left(B_{y,0}^{(f)}\right)^2, \quad f=\frac{\omega}{2 \pi}. \ee
We can assume that this spectral density is given by an analogue
filter, but in fact we have to do a Fast Fourier Transformation
(FFT) for the magnetic field as a function of time. Generally
speaking $\Phi (f)$ is a tensor, but for an illustrative purpose let
us investigate only the y-components of the Alfv\'en waves. Then the
heating power is given by a frequency integral of the damping rate
and the spectral density taken up to the maximal frequency for which
the magnetohydrodynamic approach is still applicable, i.e. the
maximal electron-proton collisions frequency
$\omega_\mathrm{max}=1/\tau_e$:
\be \label{Qap} Q_\mathrm{abs}=\frac{1}{2 \mu_0}\int_{0}^{1/2 \pi
\tau_e} \frac{\Phi(f)}{\tau_A(f)}{\rm d}f. \ee
In the damping rate we have to substitute the kinematic viscosity
with that of a hydrogen plasma \cite{LL10}
\be \nu_\mathrm{kin}=\frac{\eta}{\rho}=0.4 \frac{T_p^{5/2}}{M^{1/2}
e^4 n_p L_p}, \ee
where
\be \frac{1}{a^2}=4\pi
e^2\left(\frac{n_e}{T_e}+\frac{n_p}{T_p}\right), \quad e^2\equiv
\frac{q_e^2}{4 \pi \varepsilon_0} \ee
is the Debye radius, $T_p$ and $T_e$ are respectively the proton and
electron temperatures multiplied by the Boltzmann constant, and
\be L_p=\ln \frac{T_p a}{e^2}\gg1 \ee
is the Coulomb logarithm as the thermal velocity of the protons is
much smaller than the Bohr's velocity
\be v_{Tp}\ll v_\mathrm{Bohr}\equiv \frac{e^2}{\hbar},\quad
Mv_{Tp}^2=T_p. \ee
Here we have used the elementary gas-kinetic evaluation for the
viscosity
\be \eta \sim n_p M l_p v_{Tp}, \ee
as well as the mass density of the plasma, which is practically
determined by the density of the protons $n_p$ and their mass M
\be \rho=M n_p. \ee
In the derivation of the upper evaluation we have supposed that two
protons with kinetic energy $T_p$ can be brought to a minimal
distance $r_\mathrm{min}$
\be \frac{e^2}{r_\mathrm{min}} \sim T_p. \ee
This distance parameterizes the effective cross-section
$\varsigma_p$ for the proton-proton collisions, which participates
in the evaluation for the mean free path of the protons
\be l_p\varsigma_{p} n_p=1. \ee
The Coulomb logarithm additionally increases the cross-section
$\varsigma_p$
\be \varsigma_{p} \sim \pi (2 r_\mathrm{min})^2 L_p. \ee
For the hot coronal plasma the kinematic viscosity is much bigger
than the magnetic one $\nu_\mathrm{kin}\gg\nu_{\mathrm{mag}}.$ This
determines the diffusion of the magnetic field due to the ohmic
resistance.

For evaluating the maximal frequency at which magnetohydrodynamics
can still be applied we have to calculate the maximal frequency at
which the density of the electric current, transferred by the
electrons, $\mathbf{j}$ creates through the magnetic field
$\mathbf{B}$ bulk Lorentz force, acting on the proton fluid with
mass density $\rho$. This is in fact the frequency of the
electron-proton collisions $1/\tau_e$, which we take from the
elementary kinetic gas theory. We can use the duration of the mean
free path of electrons $\tau_e$ to determine the electric
conductivity of the plasma
\be \frac{\sigma}{4\pi \varepsilon_0} \sim \frac{e^2 n_e}{m}\tau_e.
\ee
The mean time between two electron collisions is defined by the
electron mean free path $l_e$ and their thermal velocity $v_{Te}$
\be \tau_e=\frac{l_e}{v_{Te}}. \ee
The relation between the electron density $n_e$, their free path
$l_e$ and effective cross-section $\varsigma_e$ is the same as for
protons
\be l_e n_e \varsigma_{e}=1,\quad \varsigma_{e} \sim \pi(2
r_{min})^2 L_e,  \ee
only the expression for the Coulomb logarithm should be changed
\be L_e=\ln\frac{T_e}{\hbar \omega_{pl}}\gg1, \ee
as even the velocity of the non-relativistic electrons is much
greater than the Bohr velocity
\be \frac{e^2}{\hbar}\ll v_{Te}\ll c. \ee
The plasma frequency $\omega_{pl}$ is much higher than the
collisions frequency
\be \omega_{pl}^2=\frac{n_e e^2}{m}, \ee
and yet it is located in the radio-range.

These formulas give the well-known evaluations for the
electron-proton collisions frequency
\be \frac{1}{\tau_e}=\frac{4\pi e^4 n_e L_e}{m^{1/2}T_e^{3/2}} \sim
\frac{v_{Te}}{L_e} \sim \frac{1}{\nu_e} \sim \omega_{A,c} \sim 2\pi
f_c, \ee
as well as for the conductivity of the fully ionized hydrogen plasma
\be \frac{\sigma}{4\pi\varepsilon_0}=0.6
\frac{T_e^{3/2}}{e^2m^{1/2}}. \ee
These model kinetic evaluations applied to the coronal plasma could
be used for calculation of the heating power due to absorption of
Alfv\'en waves. For an analytical illustration we will utilize the
interpolation formula, used for analysis\cite{Burlaga_Mish} of the
magnetometric data from Voyager 1
\be \Phi(f) \sim \frac{\mathcal D}{f^2}. \ee
Such a spectral density $1/f^2$ is naturally explained in the
framework of a stochastic Langevin MHD, as Mishonov and Maneva
pointed out \cite{TMYM06}, Eq.~41, and it is not related to any
phase-coherent structures which could be distinguished from
incoherent turbulent fluctuations, cf. with the work by Roberts and
Goldstein \cite{10.150}. The Kolmogorov turbulence $f^{-5/3}$ gives
\cite{10.91, 10.30} a slightly (17\%) different power. The idea that
convective granules produce a continuous stream of noise generating
waves is actually very old \cite{Schatzman}. We believe that this is
a common property of all turbulent magnetized plasmas and we suggest
the hypothesis that the $1/f^2$ spectral density could be used even
for the physics of accretion disks.

Bearing in mind all above, the general formula Eq.~(\ref{Qap}) gives
the final model evaluation for the bulk heating power
\be \label{totPower} Q_\mathrm{abs}=0.4
(2\pi)^2\frac{\mathcal D}{B_0^2}\left(\frac{M}{m}\right)^{1/2}
\frac{L_e}{L_p}\left(\frac{T_p}{T_e}\right)^{3/2}n_pT_p. \ee
This heating rate actually sets the upper limit for the absorbed
bulk power that can be derived in the magnetohydrodynamic framework
when we take into account the influence of the diffusion Alfv\'en
modes with frequencies $\omega > 1/\tau_A$ as well. However, these
diffusion modes are able to heat only the narrow ``skin layer''
above the limb. Far away from the Sun (at the height of the
satellites) we have to account only for those Alfv\'en waves,
propagating with $\omega_A > 1/\tau_A.$ If we substitute the
proton-electron collisions frequency cut-off $1/\tau_e$ in
\Eqref{Qap} with the Alfv\'en waves damping rate we will end up with
a different assessment for the heating rate
\be \tilde Q_\mathrm{abs} = \frac{\pi \mathcal D}{\mu_0}. \ee
Hereby we obtain another time constant related to kinetics of the protons
that is rather significant for the theory of coronal plasma heating
\be
\frac{1}{\tau_\mathrm{h}}=\frac{\tilde Q_\mathrm{abs}}{n_pT_p}
=\frac{\pi \mathcal D}{\mu_0n_pT_p}.
\ee
This time constant $\tau_\mathrm{h}$ together with the velocity of
the solar wind $v_\mathrm{wind}$ form a dimensionless parameter
\be
\Upsilon = \frac{r_\mathrm{Sun}}{v_\mathrm{wind}\tau_\mathrm{h}},
\ee
which shows at what distance from the Sun $r_\mathrm{Sun}$ the
discussed mechanism for coronal heating stays important and should be
incorporated in the coronal dynamics; at $\Upsilon \ll 1$ the influence of
the Alfv\'en waves heating is negligible. According to our statistical approach,
the heating time-constant depends on the choice of the frequency cut-off.
Therefore the role of the frequency cut-off for the coronal
heating due to viscous damping of magnetohydrodynamic waves is
extremely important and we wish to find out what kind of physical
processes in the plasma prevail to determine it. We have to pay
attention that the ratio of these two cut-offs significantly varies
with the solar radius, starting from a large number in the
photosphere, where the collisions frequency dominates the Alfv\'en
waves attenuation, and reaching the order of unity in the lower
corona, where the plasma $\beta$ becomes rather small
\be
\frac{\tau_{A,c}}{\tau_e}= 0.4\pi\beta
\frac{n_eL_e}{n_pL_p}{\left(\frac{MT^3_p}{mT^3_e}\right)}^{1/2}\!\!\!\!\!,
%{\left(\frac{T_p}{T_e}\right)}^{3/2}
\quad \beta = \frac{n_eT_e+n_pT_p}{\left(B_0^2/2\mu_0\right)}.
\ee
 As one can see, the most important characteristic in the result for the
absorbed power \Eqref{totPower} is the spectral density of the angle
of deviation of the magnetic field lines from their state of static
balance position.

Due to this distortion as strings of a harp
\be (b^2)_\omega=\frac{\Phi(f)}{B_0^2}=\frac{(B_{y,0}^2)_f}{B_0^2}
\approx \frac{\mathcal D}{B_0^2 f^2}, \ee
we can listen to the impressive symphony of the convective
turbulence of the Sun and its corona is heated mainly by the
high-frequency part of the spectral density.

%%%%%%%%%%%%%%%%%%%%%%%%%%%%%%%%%%%%%%%%%%%%%%%%%%%%%%%%%%%%%%%%%%%%
\section{Discussion}
%%%%%%%%%%%%%%%%%%%%%%%%%%%%%%%%%%%%%%%%%%%%%%%%%%%%%%%%%%%%%%%%%%%%

A new theory often means data mining and experimental data
processing, but our purpose is not to extract the electron and
proton temperatures from SOHO data; our purpose is to explain a
local property of the magnetized plasma, pierced by Alfv\'en waves.

We have calculated the absorption power of the Alfv\'en waves
$Q_\mathrm{abs}$, that is the power transmitted per unit volume from
the Alfv\'en waves to the protons. In order to incorporate this
leading (according to our opinion) power in a coherent picture we
ought to include $Q_\mathrm{abs}$ in the total plasma energy balance
which takes into account all possible energy losses or gains: the
work due to the gas expansion, the change of the temperature which
is proportional to the density of the kinetic energy, the energy
transmission from protons to electrons by collisions, bremsstralung
of electrons, etc. In other words, $Q_\mathrm{abs}$ should be the
last missing ingredient of an almost completed model for the coronal
heating and creation of the solar wind.

Few words should be said about how possibly our scenario could be
rejected; the main defect of the contemporary coronal investigations
is that none of the proposed mechanisms is consigned to the dustbin.
Even the magnetic reconnections which predict significant Ohmic
heating and $T_e >> T_p$. As we propose Alfv\'en waves as a heating
mechanism the perpendicular to the static magnetic field pulsations
of the velocity and the magnetic field should be in relation
corresponding to Alfv\'en waves \cite{LLproblem} Eq.~(69.11)
\be \frac{v_{\alpha}}{V_A}=-b_{\alpha}=-\frac{B_{\alpha}^{'}(z,t)}
{B_0},\quad \alpha = 1,2,3. \ee
Such a relation was observed during the first space missions
\cite{Bruno}. In other words the present heating scenario should be
abandoned if the correlation coefficient between the velocity and
the magnetic field is not high enough. According to our picture the
waves are generated by the turbulence in the dense photospheric
layer of the Sun where the influence of the magnetic field is
relatively weak and mainly transversal incompressible
magnetohydrodynamic waves (MHD) waves are generated. These waves
reach the regions with negligible pressure (i.e. low plasma $\beta$
regions) but still their polarization corresponds to MHD waves in an
incompressible fluid.

%%%%%%%%%%%%%%%%%%%%%%%%%%%%%%%%%%%%%%%%%%%%%%%%%%%%%%%%%%%%%%%%%%%%
\section{Conclusions}
%%%%%%%%%%%%%%%%%%%%%%%%%%%%%%%%%%%%%%%%%%%%%%%%%%%%%%%%%%%%%%%%%%%%
The model problem for Alfv\'en waves transmission from a magnetized
turbulent half-space to a non-turbulent half-space has been recently
solved \cite{TMYM06}. This model approach can be further developed
by an inclusion of the inhomogeneity of the magnetic field crossing
the two half-spaces interface. However, for the coronal heating
theory we consider as more appropriate to use experimental data for
the Alfv\'en waves spectral density. But the accuracy of the
magnetic field measurements is much higher than that of the
tachometry of the solar wind  - that is why we focus on
magnetometric data for the spectral density of the magnetic field
$\Phi(\omega)$. This tensor ought to have two equal eigen values and
another smaller one related to the static field component. Of
course, it would be much better to have data for the magnetic field
and velocity correlation. Investigations of the general $6\times6$
matrix, including all correlation coefficients, should be planned
for the forthcoming space missions. The most important problem is
where the spectral density of the MHD waves comes to an end. From a
microscopic point of view the cut-off frequency is at
electron-proton collisions rate $1/\tau_e$ which is well in the
range of the standard radio frequency measurements. However, the
maximal frequency $\omega_\mathrm{max}$ should be determined by the
concrete physical conditions. For example, the inverse minimal size
of the solar granulation could also determine the cut-off
wave-vector $k_c.$ Another possibility for the maximal frequency is
to be set by the waves propagation condition $\omega_A\tau_A>1$,
which results in
\be \omega_c = V_Ak_c = \nu k_c^2 = V^2_A/\nu = B^2_0/\mu_0\nu\rho,
\quad k_c=V_A/\nu. \ee
Above this frequency the waves are pure dissipative modes $\omega
\approx i\nu k^2$ with frequency-dependent depth of the ``skin
layer'' $\delta=\sqrt{\nu/\omega}.$

One of the problems we encounter today is that the high-frequency
MHD waves are absorbed during the coronal heating and all the
satellites' reports have a distorted spectral density like the red
light of a sunset. In order to observe the bare MHD waves we need to
have magnetometric data from a satellite-kamikaze, but this would be
a too expensive toy to be planned and realized. We need a more
ingenious solution. We ought to analyze online which region of the
Sun emits the MHD waves, investigated by the magnetometer.

As a real break-through of the solar physics we consider the
investigation of the correlation between the stochasticity of the
space magnetic field delivered by the MHD waves and the
stochasticity of the solar surface, obtained by optical data.
Correlation of SOHO Dopplergrams with magnetograms is an important
task, which should be put on the agenda in the global
magnetohydrodynamic modeling of the solar corona. Before this is
done we can just extrapolate the spectral density from the
experimentally accessible region to the cut-off frequency
$1/\tau_e$, as it was done in the derivation of our model evaluation
Eq.~(\ref{Qap}). This $Q_\mathrm{abs}$ can be compared with the
necessary heating power of the global MHD models for the solar
corona. This final formula contains only experimentally known
parameters and can be used in the global MHD models for the solar
corona.

We would like to close with a qualitative discussion of the proposed
mechanism in this work. As it is well-known, the leap of two orders
of magnitude of the coronal temperature is rather sharp, compared to
the height to the optically dense layers. This leap is a kind of a
domain border, typical for the condensed matter physics. This sharp
border ought to have at least qualitatively a good explanation in
each mechanism for heating of the coronal plasma and this should be
used as an important aesthetical criterion for the applicability of
each coronal heating model. For most of them the closer we are to
the dense layers the more intense is the heating and the observed
temperature jump is practically unexplainable. Let us discuss this
leap in the framework of the proposed scenario: slightly above the
optically dense layers, emitting black body radiation $\propto T^4,$
the dense plasma is transparent for the MHD waves as the viscosity
$\rho \nu \propto \eta \propto T^{5/2}$ is almost density
independent. In this optically transparent layer the heating is
negligible and the cooling is realized perhaps by a small residual
convective turbulence. Imagine that all the plasma above is cool
with $T \sim 6\;\rm{kK}$ and let us consider its temperature
stability. Imagine a local fluctuation - a small augmentation of the
temperature. This increasing will lead to the enhancement of the
viscosity $\eta$ and the heat absorption from the MHD waves,
piercing with an almost constant energy density through every layer
of the plasma. This increasing of the heating leads to an
extra-augmentation of the temperature, which cannot be compensated
form the thermal conductivity in the rarefied plasma. Thus we have a
positive feed back and the plasma explodes from the absorption of
high-frequency MHD waves. This is simultaneously the mechanism
explaining the creation of the solar wind. As a comparison let us
mention that for none of the chemical explosions at room temperature
we do not have an increasing of the temperature with as much as two
orders of magnitude. It would be of a significant interest to
investigate the leap of the spectral density of the magnetic field
$\Phi(\omega)$ at this frontier.

As a conclusion we would like to mention that in fact the first
historical explanation of the solar coronal heating problem is still
on the arena, retrieving another life or at least an $impetus$ from
the current satellites' data.

\begin{acknowledgments}
Discussion and support by D.~Damianov, R.~Erdelyi, I.~Rousev,
T.~Zaqarashvili and I.~Zhelyazkov are highly appreciated.
\end{acknowledgments}

\end{document}